\documentclass[twocolumn]{article}
\usepackage{preprint_arxiv}

\usepackage{graphicx}

\usepackage{amsmath, amsthm, amssymb, amsfonts}

\usepackage[numbers,square,sort&compress]{natbib}
\usepackage[utf8]{inputenc}	% allow utf-8 input
\usepackage[T1]{fontenc}	% use 8-bit T1 fonts
\usepackage{xcolor}		% colors for hyperlinks
\usepackage[colorlinks = true,
            linkcolor = purple,
            urlcolor  = blue,
            citecolor = cyan,
            anchorcolor = black]{hyperref}	% Color links to references, figures, etc.
\usepackage{booktabs} 		% professional-quality tables
\usepackage{nicefrac}		% compact symbols for 1/2, etc.

\usepackage{lineno}		% Line numbers
\usepackage{float}			% Allows for figures within multicol
\usepackage[labelfont=bf,labelsep=colon]{caption}
\usepackage{newfloat}
\DeclareFloatingEnvironment[name={Supplementary Figure}]{suppfigure}
\usepackage{sidecap}
\sidecaptionvpos{figure}{c}
\usepackage{microtype}		% microtypography
\usepackage{titlesec}
\titlespacing\section{0pt}{12pt plus 3pt minus 3pt}{1pt plus 1pt minus 1pt}
\titlespacing\subsection{0pt}{10pt plus 3pt minus 3pt}{1pt plus 1pt minus 1pt}
\titlespacing\subsubsection{0pt}{8pt plus 3pt minus 3pt}{1pt plus 1pt minus 1pt}

\usepackage[font={footnotesize,sf,stretch=1.0},
            labelfont={bf,sf},
            labelsep=period,
            justification=raggedright,
            singlelinecheck=false]{caption}
\title{The Demon Hidden Behind Life’s 
Ultra–Energy-Efficient Information Processing
— Demonstrated by Biological Molecular Motors
}

\usepackage{authblk}
\author[1,2,3]{Toshio Yanagida}
\author[4]{Keisuke Fujita}
\author[3,5,6]{Mitsuhiro Iwaki}

\affil[1]{Center for Information and Neural Networks (CiNet), National Institute of Information and Communications Technology (NICT), Osaka, Japan}
\affil[2]{Graduate School of Information Science and Technology, The University of Osaka; NEC Brain-Inspired Computing Collaborative Research Institute (NBIC), Osaka, Japan}
\affil[3]{Immunology Frontier Research Center (IFReC) The University of Osaka, Japan}
\affil[4]{Premium Research Institute for Human Metaverse Medicine (WPI-PRIMe), The University of Osaka, Japan}
\affil[5]{Advanced ICT Research Institute, National Institute of Information and Communications Technology (NICT), Hyogo, Japan}
\affil[6]{Graduate School of Frontier Biosciences, The University of Osaka, Japan}

\usepackage{booktabs}
\usepackage{tabularx}

\AtBeginDocument{
  \setlength{\footskip}{25pt} % ← ページ番号と本文の距離を広げる
}
\begin{document}

\twocolumn[ % Method A for two-column formatting
  \begin{@twocolumnfalse} % Method A for two-column formatting
  
\maketitle

\begin{abstract}
The remarkable progress of artificial intelligence (AI) has revealed the enormous energy demands of contemporary digital architectures, raising profound concerns about sustainability. In stark contrast, the human brain operates on only \textasciitilde20~watts, and individual cells process gigabit-scale genetic information using energy on the order of trillionths of a watt. Under an equivalent energy budget, a general-purpose digital processor can perform only a few simple operations per second, such as ``1 + 1 = 2.'' This striking disparity highlights a fundamental mystery of life, suggesting that biological systems operate according to algorithms that are intrinsically different from conventional computational principles.
The framework of information thermodynamics---most notably Maxwell's demon and the Szilard engine---offers a theoretical key to this mystery. These once-paradoxical constructs have been resolved, establishing a lower bound on the energy required for information processing. Yet, the energy consumed by general-purpose digital processors exceeds this bound by roughly six orders of magnitude. Recent single-molecule studies further show that biological molecular motors actively harness Brownian motion to convert positional information into mechanical work, thereby realizing a ``demon-like'' operational principle. Such findings strongly suggest that living systems have already implemented an ultra-efficient information--energy conversion strategy that transcends conventional computation.
Here, we experimentally quantify the physical basis of this biological information processing. Specifically, we establish an explicit correspondence between positional information (bits) and the mechanical work extracted from it, demonstrating that molecular machines selectively exploit rare but functional fluctuations---occurring approximately once in every 3{,}000 events---arising from Brownian motion on the microsecond timescale, thereby enabling millisecond-scale operation consistent with physiological ATP hydrolysis efficiency. These results indicate that information, energy, and physiological timescales are organically integrated within a single experimental framework, suggesting that life may implement a Maxwell's demon--like mechanism to achieve extreme energy efficiency in information processing.
\end{abstract}
%\keywords{First keyword \and Second keyword \and More} % (optional)
\vspace{0.35cm}

  \end{@twocolumnfalse} % Method A for two-column formatting
] % Method A for two-column formatting

%\begin{multicols}{2} % Method B for two-column formatting (doesn't play well with line numbers), comment out if using method A

%%%%%%%%%%%%%%%  Main text   %%%%%%%%%%%%%%%
% \linenumbers
\section{\textbf{Introduction}}
\begin{figure}[htbp]
\centering
\includegraphics[width=1\columnwidth]{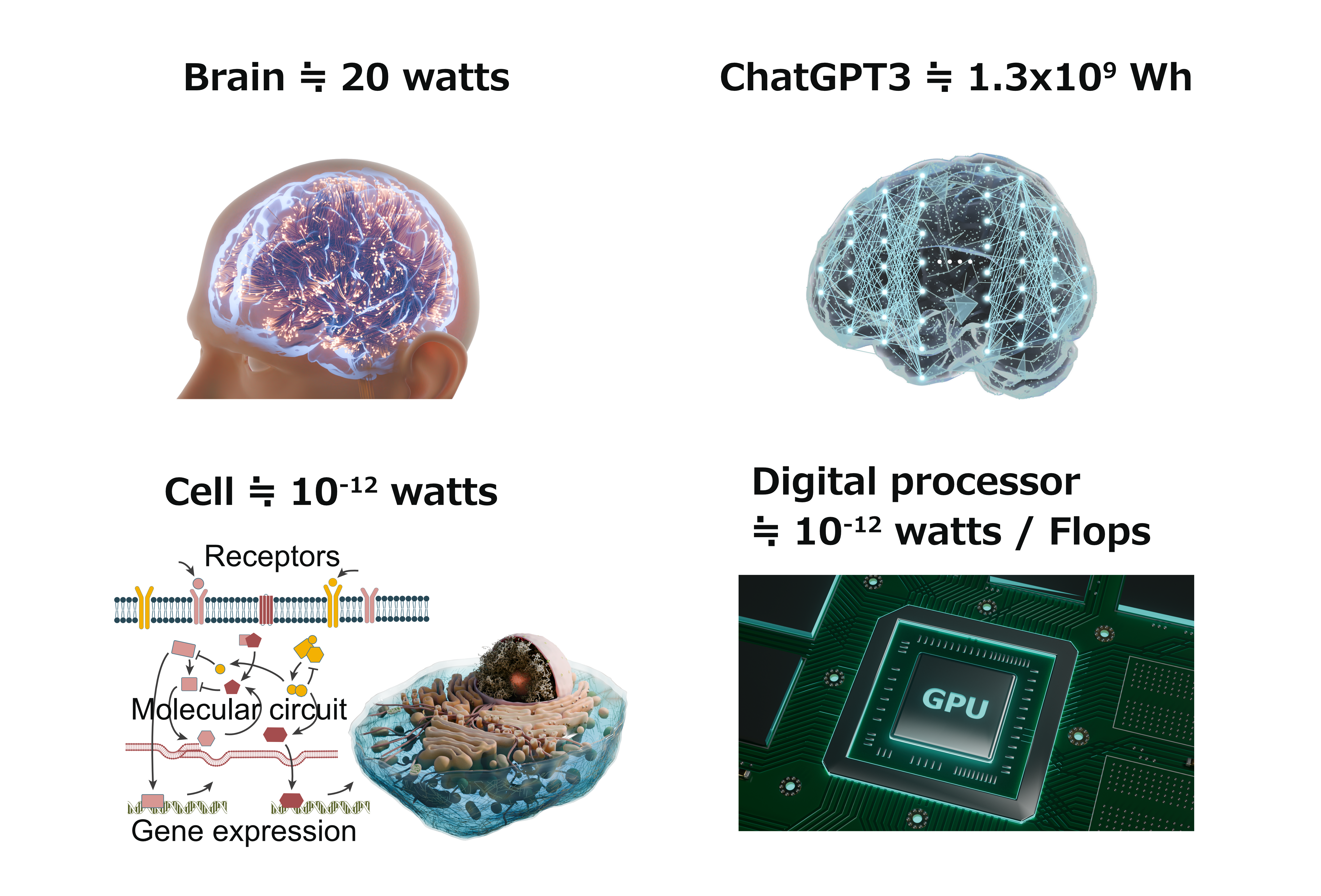}
\caption{\textbf{Comparison of energy consumption between artificial neural networks and biological brains.}}
\label{fig:fig1}
\end{figure}

The rapid development of generative AI has dramatically transformed society, but the price is enormous energy consumption. For example, training ChatGPT-3 is reported to require electricity on the gigawatt-hour scale \cite{Patterson2021}, and newer models are estimated to reach several tens of times this amount. These figures suggest that the current trajectory of AI development may be fundamentally limited by constraints of sustainability.
On the other hand, the human brain keeps its power consumption almost constant---about 20~watts---from rest to execution of thought and decision-making \cite{Raichle2006}, yet still realizes complex cognition and decision-making. Quantitative analyses indicate that most of the brain's energy is spent not on computation itself but on communication between neurons \cite{Attwell2001,Levy2021}, suggesting an information-processing mode fundamentally different from conventional computation. The contrast is even more pronounced at the cellular scale. The human body consists of $\sim 3.5 \times 10^{13}$~cells \cite{Bianconi2013}, each of which consumes only a few picowatts (pico~$=10^{-12}$) yet can access gigabit-scale genetic information through complex molecular networks \cite{Kanehisa2000,Willyard2018} and control the expression of thousands of proteins. This efficiency is incommensurable with that of a general-purpose digital processor (for example, NVIDIA H100), which can perform only a few simple operations per second with the same amount of energy. Such a gap strongly suggests that unknown principles of information--energy conversion are actually at work in biological systems (Fig.~\ref{fig:fig1}).

A promising framework for understanding this is information thermodynamics, exemplified by Maxwell’s demon \cite{Maxwell1871} and the Szilard engine \cite{Szilard1929}. Once regarded as paradoxes, these thought experiments are now being recognized as embodying ultra-efficient algorithms for information processing \cite{Bennett1982,Sagawa2012,SagawaUeda2013,Toyabe2010,Berut2012,Koski2014,Chida2017}. At the same time, advances in state-of-the-art single-molecule imaging and manipulation \cite{Funatsu1995,Finer1994,Ishijima1998,Yildiz2003,Vale1996,Noji1997} are clarifying mechanisms by which biological molecular motors bias probabilistic dynamics using Brownian motion and convert positional information into mechanical energy \cite{Kitamura1999,Kitamura2005,Fujita2019,Dunn2007,Shiroguchi2007,Iwaki2009,Nishikawa2010,Fujita2012,Okada1999,Carter2005,ReckPeterson2006,Watanabe2013,Astumian1997,Julicher1997}.

Contributions. This work goes beyond prior ratchet-based descriptions by (i) quantitatively linking positional information $I$ (bits) to extractable work $W = k_{\mathrm{B}} T \ln 2 \times I$, (ii) bridging timescales by showing how $\sim 100~\mu\mathrm{s}$ Brownian fluctuations selected with probability $W_i \approx 1/3000$ yield a mean ATP turnover of $\tau \approx 300~\mathrm{ms}$ at real physiological efficiency, and (iii) extending the information--energy principle revealed in molecular motors toward design rules for noise-driven AI architectures.

\section{\textbf{Maxwell’s demon and energy-efficient information processing}}
We formalize the information$\rightarrow$work lower bound and notation (e.g., $k_{\mathrm{B}} T \ln 2$) used later for quantitative comparisons.

\begin{figure}[htbp]
  \centering
  \includegraphics[width=0.8\linewidth]{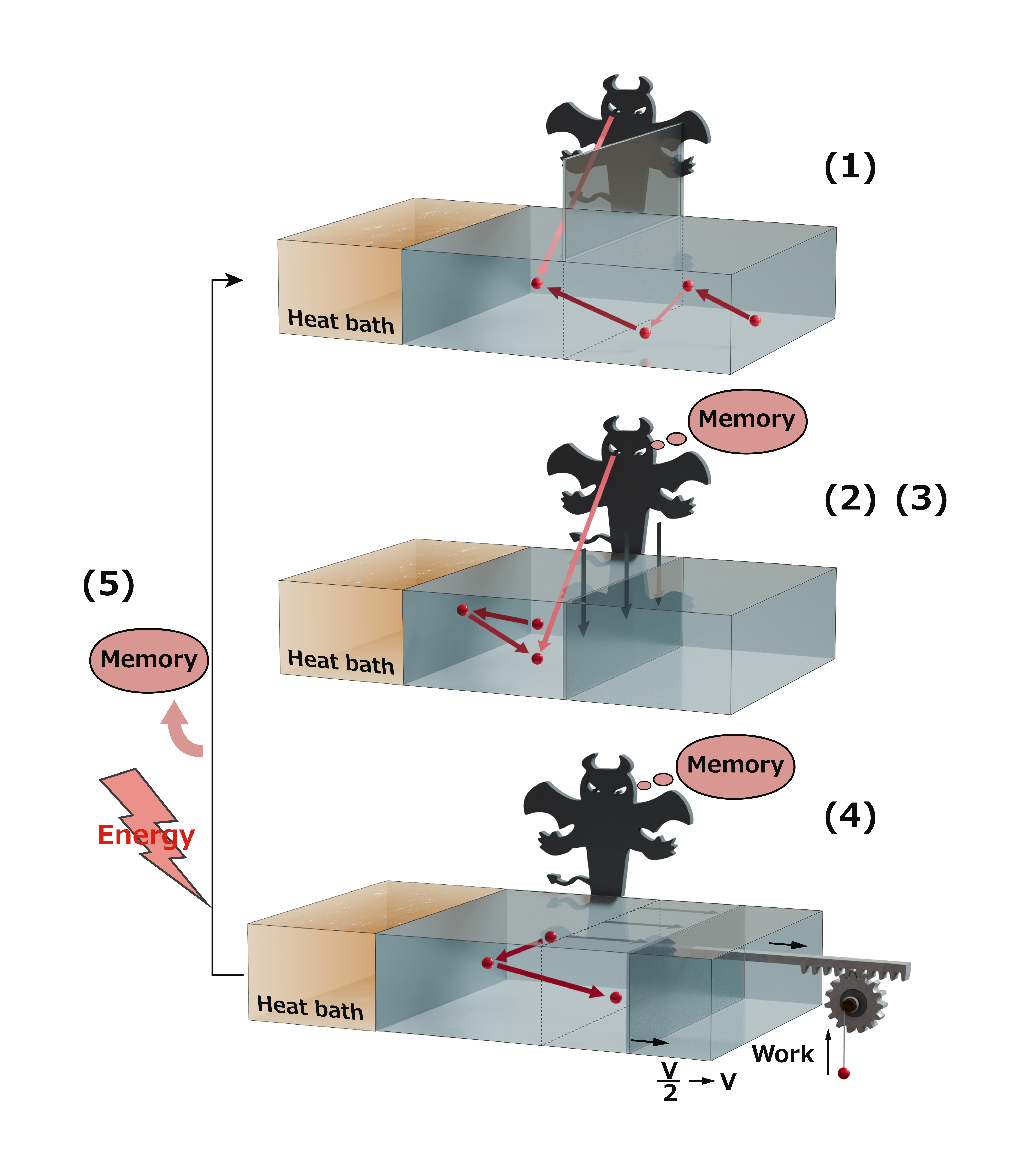} % ←画像ファイル名に置き換え
  \caption{\textbf{Szilard engine and Maxwell's demon.}
  The demon observes a single molecule undergoing Brownian motion in a space connected to a heat bath (1).
  When the molecule is detected to have randomly moved to the right side of the space (2),
  the demon acquires and stores this information (1 bit) and then inserts a partition (piston) at the center (3).
  The partition is pushed leftward by collisions with the molecule,
  and the molecule expands isothermally into the available space.
  Through this isothermal expansion, the system performs work of $k_{\mathrm{B}}T\ln 2$ on the surroundings (4).
  Subsequently, the demon erases its memory by consuming $k_{\mathrm{B}}T\ln 2$ of energy,
  thereby returning the system to its initial state (5).
  }
  \label{fig:fig2}
\end{figure}

In 1871, J.~C.~Maxwell proposed the thought experiment of a ``demon,'' challenging the second law of thermodynamics (the law of entropy increase) \cite{Maxwell1871}. Several decades later, L.~Szilard abstracted this to a single-molecule model and formulated it as the ``Szilard engine'' \cite{Szilard1929}. In this model (Fig.~\ref{fig:fig2}), a demon observes a molecule undergoing Brownian motion in a closed container and inserts a partition according to its position, thereby rectifying probabilistic fluctuations and extracting work to the outside. The extractable work is expressed as
\[
W = k_{\mathrm{B}} T \ln 2
\]
Here, one bit of the molecule’s positional information (``left'' or ``right'') corresponds to extractable energy $k_{\mathrm{B}} T \ln 2$. Then at which stage does the demon consume energy? Early interpretations claimed that observation or measurement itself requires energy \cite{Brillouin1951}, but C.~H.~Bennett showed that measurement and storage of memory are, in principle, reversible and do not involve energy dissipation \cite{Bennett1982}. The unavoidable cost $\Delta E$ arises when erasing memory to reset the system, and this is formulated by Landauer’s principle as
\[
\Delta E = k_{\mathrm{B}} T \ln 2
\]
\cite{Landauer1961}. At room temperature (about 300~K), this amounts to only $\sim 3 \times 10^{-21}~\mathrm{J}$ per bit. By contrast, general-purpose CMOS logic elements consume about $10^{-15}~\mathrm{J}$ per bit operation, a difference of five to six orders of magnitude. This ``information--energy equivalence'' provides a framework that unifies thermodynamics and information theory \cite{Bennett1982,Sagawa2012,SagawaUeda2013} and offers a theoretical foundation for the extreme energy efficiency exhibited by biological systems. The implications are profound. If life implements an algorithm akin to Maxwell’s demon at the molecular level, then biological systems already embody an information--energy conversion strategy unattainable for digital machines. This possibility motivates direct experimental verification in intracellular molecular machines to determine whether such ``demon-type'' processes actually operate.

\section{\textbf{Is the demon working inside living systems?}}
We identify where positional information is acquired and how it is converted into work by molecular motors at the single-molecule level.

The central question is whether the ``mechanism for converting information into energy'' in Maxwell’s demon is actually implemented within biological molecular machines. To probe this, we focus on protein molecules—molecular motors—that convert chemical energy into mechanical motion.
\begin{figure}[htbp]
  \centering
  \includegraphics[width=0.8\linewidth]{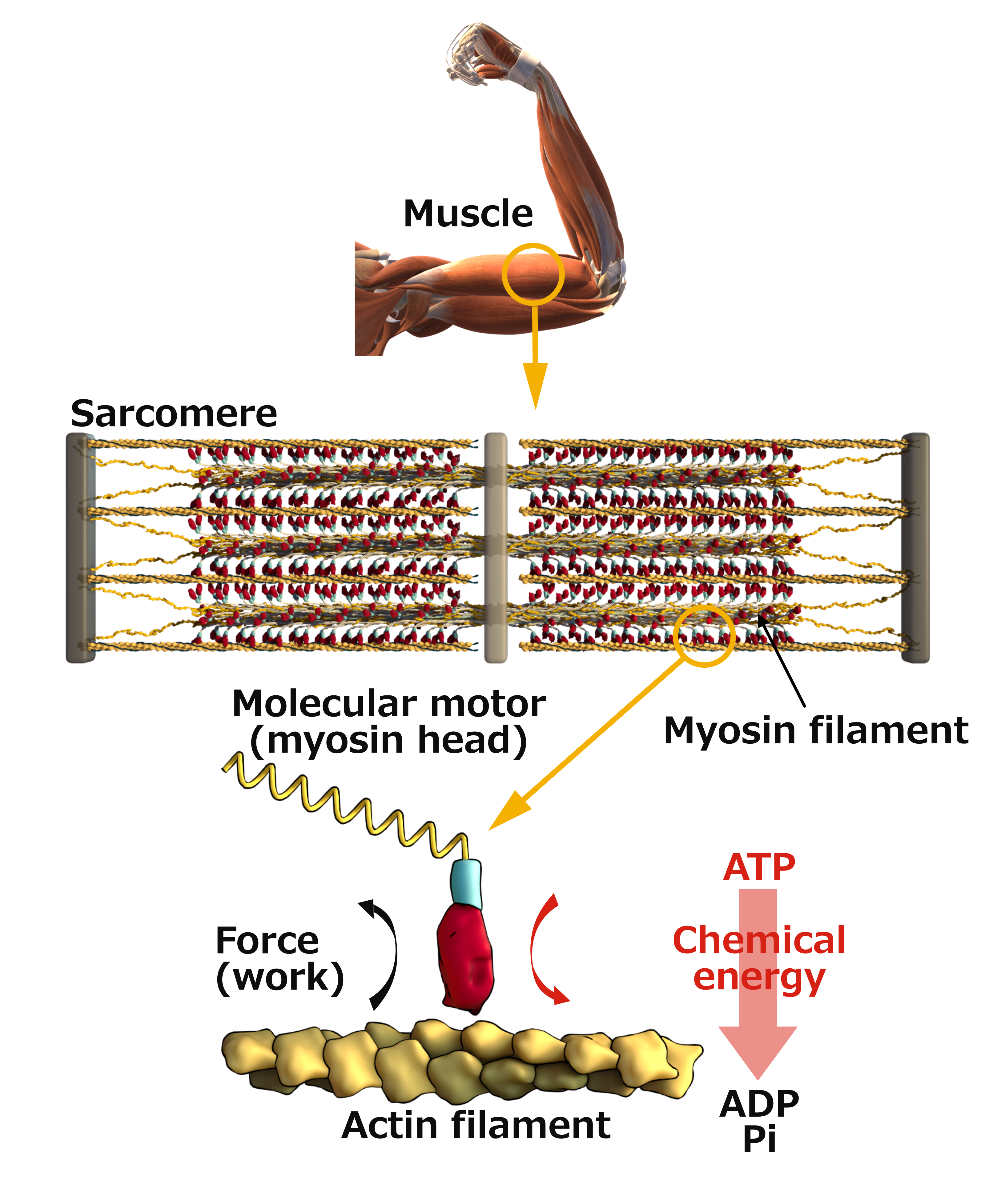} % ←画像ファイル名に置き換え
  \caption{\textbf{Molecular motors in muscle.}
  }
  \label{fig:fig3}
\end{figure}
\begin{figure}[htbp]
  \centering
  \includegraphics[width=0.8\linewidth]{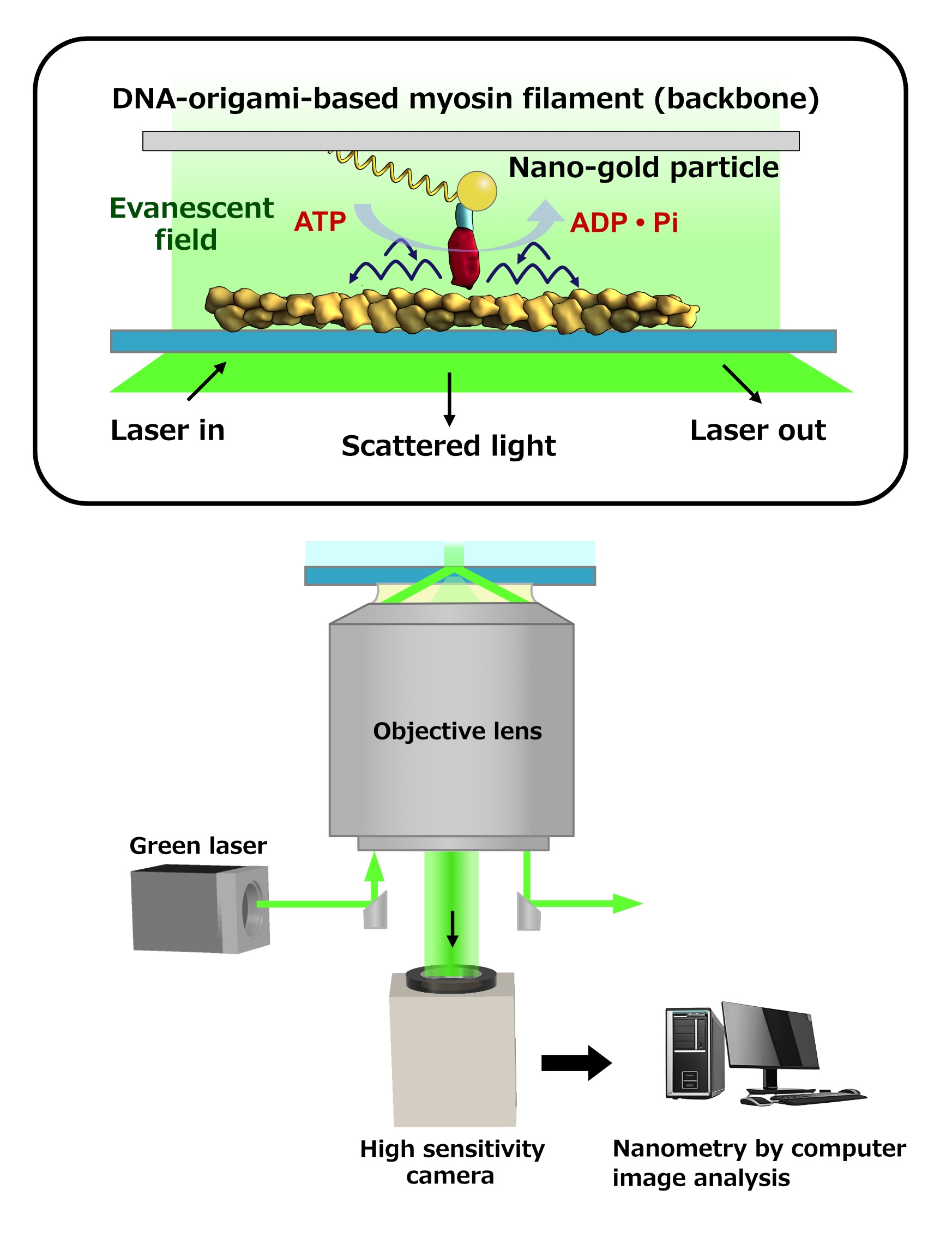} % ←画像ファイル名に置き換え
  \caption{\textbf{High-resolution single-molecule measurement of the molecular motor myosin.}
  A myosin head (molecular motor) protruding via an elastic element (spring) from a DNA-origami-based myosin filament backbone is allowed to interact with an actin filament immobilized on a glass surface in the presence of ATP. The dynamics of the myosin head are observed by tracking the position of a gold nanoparticle selectively attached to it. The gold nanoparticle is illuminated by the evanescent field generated through total internal reflection of a green laser at the glass surface (TIRF illumination), and the resulting scattered light image is captured with a high-speed camera. Image analysis enables measurement of the centroid position of the gold nanoparticle—and thereby of the molecular motor—with a spatial resolution of $0.7~\mathrm{nm}$ and a temporal resolution of $40~\mu\mathrm{s}$ \cite{Fujita2019}.
  }
  \label{fig:fig4}
\end{figure}
A representative molecular motor is muscle myosin. Myosin interacts with actin to generate the contractile force of skeletal muscle (Fig.~\ref{fig:fig3}). Muscle contraction occurs through the mutual sliding of actin and myosin filaments, and the driving force is supplied by myosin heads protruding from the myosin filament backbone via elastic elements \cite{Kaya2010}. Each myosin head is an ellipse of approximately $20~\mathrm{nm} \times 10~\mathrm{nm}$, equipped with both an ATP hydrolysis site and an actin-binding site \cite{Rayment1993}. The core issue here is: ``How does myosin convert the chemical energy of ATP hydrolysis into mechanical work?'' Historically, this problem has been investigated using physiological and biochemical experiments with muscle fibers and purified proteins \cite{Goldman1987}. In the late 1980s, it became possible to directly observe and manipulate single actin filaments, reconstructing and measuring motion \textit{in~vitro} \cite{Yanagida1984,Kron1986,Kishino1988,Ishijima1991}. Subsequently, methods were developed to measure, with high resolution, the mechanical response and chemical cycle of a single myosin molecule \cite{Funatsu1995,Finer1994,Ishijima1998,Kitamura1999}, greatly advancing our understanding of motor mechanisms \cite{Karagiannis2014,Spudich2024}.
Furthermore, we constructed artificial myosin thick filaments using DNA origami as a scaffold. The design allowed myosin heads to protrude via elastic springs; gold nanoparticle probes were covalently attached at positions that do not impair function, and interactions with actin filaments immobilized on glass surfaces were observed (Fig.~\ref{fig:fig4}) \cite{Fujita2019}. Tracking gold nanoparticle motion with high-sensitivity, high-speed detectors at $40~\mu\mathrm{s}$ temporal and $0.7~\mathrm{nm}$ spatial precision revealed that weakly bound myosin heads perform random walks back and forth along actin. Only when a head happens to diffuse forward does it bind strongly to an actin site ahead; motion then stops there and the internal spring is stretched. Releasing this tension generates force. This ``forward-selective binding'' is likely mediated by a strain-sensing mechanism inherent in the myosin head \cite{Iwaki2009}. In other words, the myosin motor, like Maxwell’s demon, acquires positional information from Brownian motion and converts it into mechanical work.

\begin{figure}[htbp]
\centering
\includegraphics[width=0.8\linewidth]{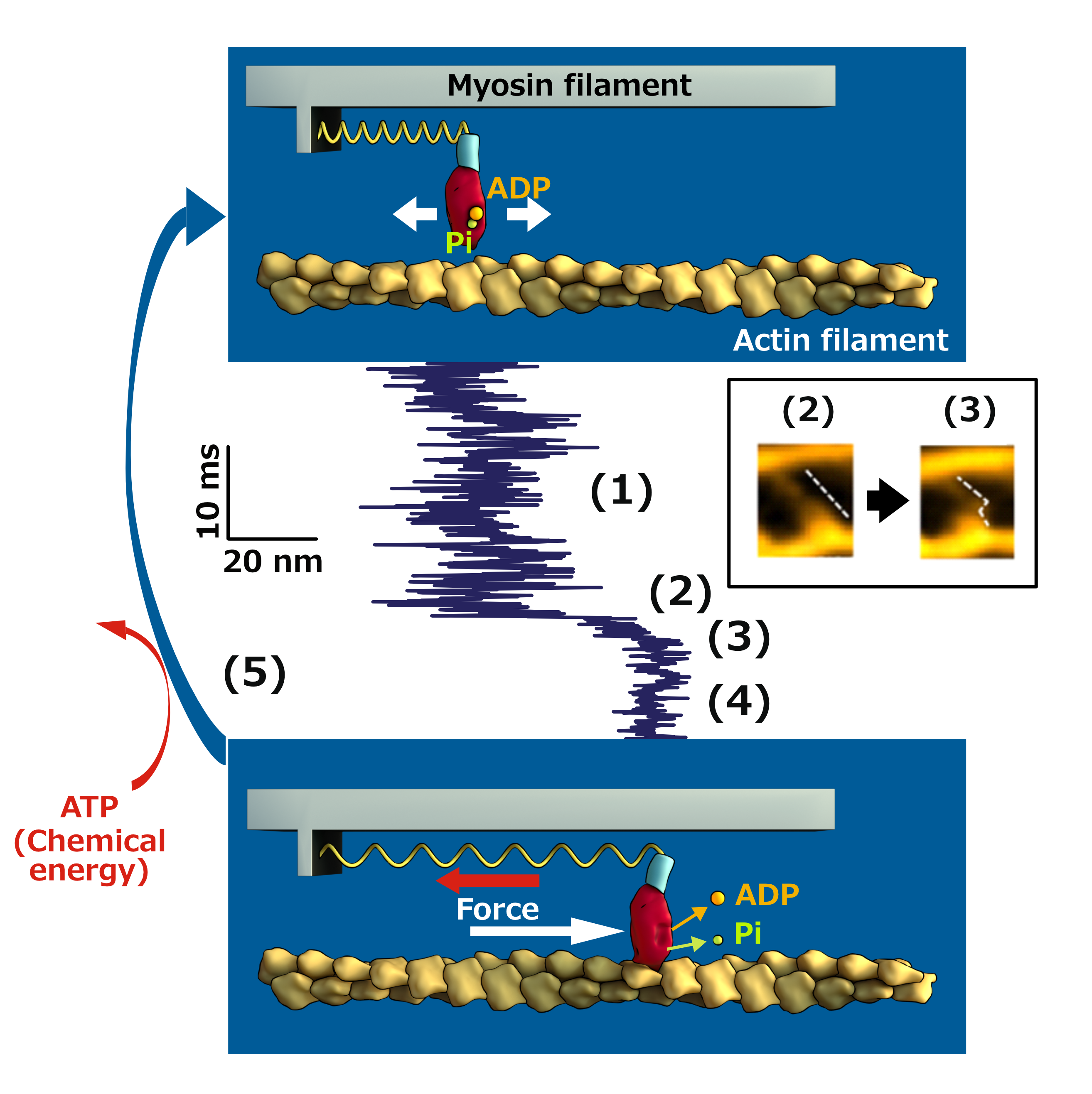}
\caption{\textbf{Molecular motor converting positional information into motion by exploiting Brownian fluctuations.}\ 
The dynamics of a molecular motor measured by high-resolution single-molecule analysis (Fig.~\ref{fig:fig4}) are shown. 
The central trace represents the actual trajectory of the observed molecular motor. 
The motor repeatedly binds to and dissociates from the actin filaments on a microsecond timescale, undergoing forward and backward Brownian motion. 
When it happens to fluctuate forward far enough to contact the next actin subunit, the strain sensor of the motor is activated and binding occurs~(2). 
Upon release of ADP and Pi generated by ATP hydrolysis, the motor undergoes a structural transition and binds strongly and stably to the forward actin subunit~(3). 
As it moves, the elastic element (spring) connecting the extended myosin head to the myosin filament generates force between the two filaments~(4). 
When a new ATP molecule (energy carrier) binds to the motor, the strong actin attachment is released, and Brownian motion resumes~(5). 
The circled numbers correspond to the processes illustrated in Fig.~\ref{fig:fig2}. 
The third panel (right) shows part of a high-speed AFM movie directly capturing the structural change (tail bending) of the myosin head during step from~(2)~to~(3)~\cite{Fujita2019}.}
\label{fig:fig5}
\end{figure}
Table \ref{tab:demon_motor} summarizes the thermodynamic correspondence between the Szilard-engine-type demon and biological molecular motors (myosin). Thus, molecular motors are not merely “similar” to demons; rather, they embody an operationally equivalent process—conversion from information to energy—that can be experimentally observed at the molecular scale.
\begin{table}[t]
  \caption{Thermodynamic correspondence between Maxwell’s demon and the biological molecular motor (myosin)}
  \centering
  \renewcommand{\arraystretch}{1.2} % 行間を少し広げる
  \begin{tabularx}{\columnwidth}{>{\raggedright\arraybackslash}p{2.8cm}X X}
    \toprule
    Element & Maxwell’s demon & Molecular motor (myosin) \\
    \midrule
    Active particle & Single molecule undergoing Brownian motion in a box & Single myosin head undergoing Brownian motion along actin \\
    Information & Measurement of molecular position (left/right) & Detection of forward position via strain-sensing mechanism \\
    Memory & Storage of positional information & Retention of position via strong actin binding \\
    Action & Insertion of partition to confine the molecule & Elastic stretching between myosin head and actin filament \\
    Work output & Molecule pushes the partition to perform work & Elastic recoil generates external mechanical force \\
    Energy dissipation & Energy cost of erasing memory ($\Delta E = k_{\mathrm{B}} T \ln 2$) & Dissociation of strong actin--myosin binding (ATP binding) \\
    \bottomrule
  \end{tabularx}
  \label{tab:demon_motor}
\end{table}

\section{\textbf{Verification of the “biological demon” at physiological timescales}}
We close the timescale loop, connecting $\sim100~\mu\mathrm{s}$ fluctuation selection (probability $W_i$) to the observed $\sim300~\mathrm{ms}$ ATP cycle and to the effective information $I \approx 11~\mathrm{bits}$ per cycle, thereby achieving physiological efficiency in energy conversion.

Single-molecule observations indicate that molecular motors such as myosin possess a “demonic” algorithm that extracts information from Brownian motion and converts it into mechanical energy. A natural question then arises: does this information processing function effectively at actual physiological timescales?
The archetypes of Maxwell’s demon and the Szilard engine presuppose quasi-static---i.e., extremely slow, idealized---processes. By contrast, biological systems must accomplish tasks under strict temporal constraints. To connect theory with biological reality, we examined the relevant timescales.
The myosin head undergoes Brownian motion along actin filaments on the $100$--$\mu\mathrm{s}$ timescale~\cite{Fujita2019}. If myosin converts a binary choice of ``forward/non-forward''—that is, one bit of information—into work (energy), then the extractable work per ATP hydrolysis would be
\begin{equation*}
\Delta E = k_{\mathrm{B}} T \ln 2 \approx 0.7\,k_{\mathrm{B}} T
\end{equation*}
Meanwhile, the free energy released by hydrolysis of one ATP molecule is about $20\,k_{\mathrm{B}}T$. This calculation would mean that only $\sim$3.5\% of ATP energy is utilized, far below the $\sim$40\% efficiency observed in real muscle near maximum efficiency~\cite{Marcucci2012}. The key to resolving this discrepancy lies in energy fluctuations of Brownian motion. While the average thermal energy is $1\,k_{\mathrm{B}}T$, occasionally a large fluctuation occurs—on the order of $8\,k_{\mathrm{B}}T$, corresponding to 40\% of ATP energy ($20\,k_{\mathrm{B}}T \times 0.4 = 8\,k_{\mathrm{B}}T$). Its probability $W_i$ is given by the Boltzmann distribution as
\begin{equation*}
W_i = \exp\!\left(-\frac{\Delta E}{k_{\mathrm{B}}T}\right)
     = \exp\!\left(-\frac{8\,k_{\mathrm{B}}T}{k_{\mathrm{B}}T}\right)
     \approx \frac{1}{3000}.
\end{equation*}
Therefore, about once in 3000 fluctuations on the $\sim100$--$\mu\mathrm{s}$ scale, the myosin head can exploit a high-energy fluctuation of $\sim8\,k_{\mathrm{B}}T$, yielding an efficiency of 40\%. The average ATP turnover time $\tau$ then becomes
\begin{equation*}
\tau = (100\,\mu\mathrm{s}) \times 3000 = 300\,\mathrm{ms}
\end{equation*}
This time matches well the ATP hydrolysis cycle time of myosin working near maximal efficiency under load in muscle~\cite{Marcucci2012}. Thus, there is strong support that this “demonic selection process” operates on biologically reasonable timescales.
Moreover, by selectively capturing such rare high-energy fluctuations, the myosin head effectively processes a large amount of information per cycle. Using Shannon’s information theory~\cite{Shannon1948}, the information content $I$ (in bits) for capturing an event with probability $W_i = \frac{1}{3000}$ is
\begin{equation*}
I = -\log_{2} W_i \approx 11~\mathrm{bits}.
\end{equation*}
That is, in each cycle, myosin uses about 11 bits’ worth of information to realize directional steps and force generation. In other words, myosin acts as a “temporal filter,” extracting rare high-energy states from microsecond-scale fluctuations and integrating them into coordinated mechanical output on the millisecond scale.

The apparent slowness of biological responses is not a limitation but an active strategy for extracting optimal states from stochastic fluctuations. Although slowness is often viewed as inefficiency in computers, in living systems it constitutes a positive feature that refines information selection and ensures adaptive functionality.

\section{\textbf{Mechanisms of muscle contraction: Brownian motion and structural change}}
We integrate the structural change and Brownian rectification mechanisms and discuss how each contributes to their respective roles.

The mechanism of force generation in muscle has long been studied, and two major models have been proposed. The first is the swinging cross-bridge model, independently proposed in the 1970s by H.~E.~Huxley~\cite{Huxley1969} and A.~F.~Huxley~\cite{Huxley1971}. In this model, ATP hydrolysis drives structural changes in the myosin head, and “swing” or tilting motion pulls on actin to generate force. Based on structural biology, this model that directly links protein structure to mechanical output came to be widely accepted.
However, since the late 1980s, molecular-probe studies have revealed discrepancies: the angular displacement of labeled myosin heads during contraction did not match predictions of the swing model~\cite{Yanagida1981,Cooke1982}. In response, re-examination led to the lever-arm hypothesis~\cite{Spudich2024,Uyeda1996}. This proposes that the orientation of the myosin head changes little; rather, a small change in one part (the converter domain) is amplified by rotation of the attached lever arm to produce movement. This interpretation is now supported by extensive experimental evidence~\cite{Karagiannis2014,Spudich2024,Sweeney2010}.
The second is the Brownian motion model, a theoretical hypothesis proposed in 1957 by A.~F.~Huxley~\cite{Huxley1957}. In this model, the myosin head performs thermally driven random forward–backward motion along actin, and forward movements are selectively captured to produce force. Although initially lacking direct experimental support, our single-molecule studies~\cite{Kitamura1999,Kitamura2005,Fujita2019} provided quantitative evidence for this “Brownian search-and-catch” mechanism and clarified molecular details of each reaction step.
In our recent work, we directly observed both structural changes (lever-arm swing) and Brownian motion simultaneously (Fig.~\ref{fig:fig5})~\cite{Fujita2019}. For example, in myosin~V, which transports intracellular organelles, about 20–30\% of the total mechanical output derives from lever-arm rotation, while the remaining 70–80\% comes from rectified Brownian motion~\cite{Fujita2012}.
These findings suggest that structural changes function less as the primary source of propulsion and more as control elements. Specifically, they:
\begin{itemize}
  \item tune the timing of binding events (select meaningful events from probabilistic fluctuations).
  \item maintain a positional “memory” through strong actin binding.
  \item reset states by dissociating strong binding upon ATP binding.
\end{itemize}
From this perspective, structural changes guarantee selectivity and adaptability in probabilistic environments and support the flexibility—an essential feature of biological actuators—of muscle. In contrast to artificial actuators that rely on deterministic motion, biological molecular motors incorporate environmental fluctuations into their control architecture. Therefore, the mechanism of force generation in muscle should be understood not as a simple dichotomy of “swinging lever arm” vs “Brownian motion,” but as a hybrid mechanism that complementarily integrates both~\cite{Fujita2019,Fujita2012,Hwang2019,Nie2014,Guhathakurta2015,Fujii2017}.

\section{\textbf{Extension to biological molecular machines in general}}
We generalize the information--energy mapping and fluctuation selection beyond myosin to other motors and enzymes, highlighting common Maxwell’s demon-like features.

The foregoing discussion has focused mainly on the information--energy conversion mechanism of skeletal muscle myosin, but the underlying principle is likely not unique to muscle. Rather, it may represent a universal design principle common to diverse molecular machines widely present in living organisms. Indeed, single-molecule studies of myosin~V and myosin~VI, which are responsible for intracellular transport, have shown that these motors realize directional motion by rectifying Brownian motion~\cite{Dunn2007,Shiroguchi2007,Iwaki2009,Nishikawa2010,Fujita2012}. Similarly, for microtubule-based motors such as kinesin and dynein~\cite{Okada1999,Carter2005,ReckPeterson2006}, and even for the ATP synthase $F_{1}$~\cite{Watanabe2013}, reports support the existence of Brownian-ratchet-type mechanisms that exploit thermal fluctuations for forward steps.
The use of thermal noise is not limited to molecular motors; it has also been observed in Ca$^{2+}$-ATPase~\cite{Toyoshima2004}, RNA polymerase during transcription~\cite{Abbondanzieri2005,Hodges2009}, and even in cell-signaling proteins~\cite{Kobilka2012}. This principle has inspired engineering as well, where nanoscale artificial molecular machines achieve motion using Brownian-ratchet designs; the field was recognized by the 2016 Nobel Prize in Chemistry~\cite{Feringa2017}. On the theoretical side, the operation of biological molecular machines has long been analyzed within the framework of Feynman’s ratchet~\cite{Feynman1963,Vale1990}. Further developments include numerous information-thermodynamic models describing the process of converting Brownian motion into mechanical work~\cite{Astumian1997,Julicher1997,Cordova1992,Sekimoto1997,Takano2010,Ariga2021}.
At the macroscopic scale, as in muscle contraction, large numbers of myosin molecules are regularly arranged and act cooperatively. These assemblies are likely to function not merely as collections of independent force generators, but rather as parallel, distributed information-processing systems~\cite{Kaya2017,Fukunaga2023}. Their cooperative action is probably integrated by “demonic algorithms” that utilize thermal fluctuations for decision-making and control.
In sum, the evidence strongly suggests that ultra-energy-efficient, Brownian-motion-driven information processing based on Maxwell’s demon is not a peculiar feature confined to specific molecular motors but a general and universal principle permeating biological molecular machines. Cells perform advanced information-processing tasks through complex nanoscale molecular networks, yet their power consumption is only a few picowatts per cell. This astonishing energy efficiency is likely rooted in molecular components operating under “demonic algorithms” that actively exploit thermal fluctuations. From a kinetic viewpoint, many cellular processes are driven by ATP hydrolysis with millisecond-scale reaction times, while the underlying Brownian motion occurs on the microsecond scale. This hierarchical relationship between the timescale of fluctuations and ATP utilization suggests deep structural and dynamical commonalities among molecular motors and other molecular machines. Overall, these lines of evidence strongly support that “demonic algorithms” truly operate at the scale of biological molecules.
\section{\textbf{Toward energy-efficient AI inspired by biology}}
We extract design principles (noise as a computational resource; reservoir dynamics) and offer insights for low-energy AI architectures.

The information--energy conversion algorithm based on Brownian motion elucidated in this study achieves energy efficiency far surpassing that of conventional digital computing architectures. Importantly, this principle is not confined to the molecular scale; in theory, it can be extended to the level of neural networks. Neuroscience offers a salient precedent: the brain’s energy consumption changes little between rest and task execution~\cite{Raichle2006}, suggesting that thought and decision-making arise not from additional metabolic cost but from a process of selecting meaningful patterns from ongoing spontaneous activity~\cite{Murata2014,Murata2021}. This principle of “selection from fluctuations” closely resembles the mechanism by which molecular motors rectify Brownian motion to produce forward steps. In both cases, useful output is generated by extracting structured information from probabilistic dynamics and biasing the system toward functional states.
At the neural level, fluctuations are not mere thermal noise but exist within chaos-like dynamics structured by recurrent networks~\cite{Skarda1990,Tsuda2001,Kaneko2018}. Such structured chaos amplifies local perturbations in cascades, enabling the brain to make rapid, adaptive decisions despite individual neurons being inherently slower than silicon processors.
From an engineering perspective, these insights suggest a paradigm shift toward ultra-low-power AI: rather than relying on clock-driven deterministic operations, maintain a “reservoir” of stochastic dynamics and harvest fluctuations within it for computation. Actively using noise as a computational resource could reduce the energy cost per bit of information processing by orders of magnitude. In fact, Boltzmann machines and recurrent neural networks that exploit chaotic dynamics already embody aspects of this principle~\cite{Ackley1985,Jaeger2004,Kurikawa2023}.
We propose that the “demon-type” algorithms implemented in biological molecular motors and neural systems can serve as blueprints for scalable and sustainable information-processing technologies—from nanoscale artificial machines to brain-like AI. These designs hold the potential to approach the ultimate energy efficiency demonstrated by life.

\section{\textbf{Acknowledgements}}
We thank Prof. Kunihiko Kaneko (Niels Bohr Institute), Dr. Kazushi Hosoda (NiCT, CiNet) and Dr. Tsutomu Murata (NiCT, CiNet) for valuable comments and discussions throughout the paper, particularly on the effectiveness of chaotic dynamics related to the development of energy-efficient AI.
%%%%%%%%%%%% Supplementary Methods %%%%%%%%%%%%
%\footnotesize
%\section*{Methods}

%%%%%%%%%%%%% Acknowledgements %%%%%%%%%%%%%
%\footnotesize
%\section*{Acknowledgements}

%%%%%%%%%%%%%%   Bibliography   %%%%%%%%%%%%%%
\normalsize
\bibliography{references}

%%%%%%%%%%%%  Supplementary Figures  %%%%%%%%%%%%
%\clearpage

%%%%%%%%%%%%%%%%   End   %%%%%%%%%%%%%%%%
%\end{multicols}  % Method B for two-column formatting (doesn't play well with line numbers), comment out if using method A
\end{document}